\documentclass[10pt, a4paper, twocolumn, twoside]{article} 

\usepackage[english]{babel} 

\usepackage{microtype} 

\usepackage{amsmath,amsfonts,amsthm} 

\usepackage[svgnames]{xcolor} 

\usepackage[small, labelfont=bf, up]{caption} 

\usepackage{booktabs} 

\usepackage{lastpage} 

\usepackage{graphicx} 

\usepackage{enumitem} 
\setlist{noitemsep} 

\usepackage{sectsty} 
\allsectionsfont{\usefont{OT1}{phv}{b}{n}} 

\usepackage{geometry} 

\usepackage[T1]{fontenc} 
\usepackage[utf8]{inputenc} 

\usepackage{XCharter} 

\usepackage{hyperref}

\usepackage{journals}

\usepackage{fancyhdr} 
\newcommand{\shorttitle}[1]{\fancyhead[CE]{\textsl{#1}}}
\newcommand{\shortauthors}[1]{\fancyhead[CO]{\textsl{#1}}}

\fancypagestyle{firstpage}{ 
	\fancyhf{}
        \lhead{\vspace*{0.0em} \textit{\footnotesize 21$^{\rm st}$ European Workshop on White
            Dwarfs \\ Castanheira, Campos, Montgomery, eds. \\[-0.2em]
          July 23--27, 2018, Austin, Texas, USA}}
}

\date{}

\newcommand{\authorstyle}[1]{{\large\usefont{OT1}{phv}{b}{n}\color{DarkRed}#1}} 

\newcommand{\institution}[1]{{\footnotesize\usefont{OT1}{phv}{m}{sl}\color{Black}#1}} 

\usepackage{titling} 

\newcommand{\HorRule}{\color{DarkGoldenrod}\rule{\linewidth}{1pt}} 

\pretitle{
	\vspace{-30pt} 
	\HorRule\vspace{10pt} 
	\fontsize{16}{12}\usefont{OT1}{phv}{b}{n}\selectfont 
	\color{DarkRed} 
}

\posttitle{\par\vskip 10pt} 

\preauthor{} 

\postauthor{ 
	\vspace{0pt} 
	{\par\HorRule} 
	\vspace{-10pt} 
}

\newcommand{\newabstract}[1]{
    {\section*{Abstract}
    \bfseries #1}
  }

\usepackage[authoryear]{natbib}
\title{Dusty Exoplanetary Debris Disks in the Single-Temperature Blackbody Plane} 

\shorttitle{Dusty Exoplanetary Debris Disks in the Single-Temperature Blackbody Plane} 

\shortauthors{Dennihy, Clemens, and Dunlap} 

\author{
        \authorstyle{E.~Dennihy,$^{1,2}$ J.~C.~Clemens,$^2$ and B.~H.~Dunlap$^2$}
	\newline\newline 
	$^1$\institution{Gemini Observatory, Casilla 603, La Serena, Chile; 
          edennihy@gemini.edu}\\ 
	$^2$\institution{University of North Carolina at Chapel Hill, Chapel Hill, NC, USA} 
      }

\begin{document}

\maketitle 

\thispagestyle{firstpage} 


\newabstract{
  We present a bulk sample analysis of the metal polluted white dwarfs which also host infrared bright dusty
debris disks, known to be direct signatures of an active exoplanetary accretion source. We
explore the relative positions of these systems in a “single-temperature
blackbody plane”, defined as the temperature and radius of a single-temperature blackbody
as fitted to the infrared excess. We find that the handful of dust systems which also host
gaseous debris in emission congregate along the high temperature boundary of the dust disk
region in the single-temperature blackbody plane. We discuss interpretations of this boundary and propose the single-temperature blackbody plane selction technique for use in future targeted searches of gaseous emission.
  }


\section{Background and Motivation}

The potential of white dwarfs to measure the internal chemical compositions of rocky exoplanets has recently begun to be realized \citep{jur14:areps42}. The key to unlocking this potential is an understanding of the delivery and diffusion of the exoplanetary material onto and below the white dwarf surface. Generally speaking, the delivery of these exoplanetary remnants to the white dwarf surface occurs through the scattering of smaller, rocky bodies to within the tidal disruption radius of the white dwarf. The disrupted material settles into a compact accretion disk around the white dwarf star, depositing material to the surface via a variety of accretion disk phenomena. Once on the surface, the accreted material diffuses through the observable atmosphere where its relative abundances can be measured and compared against known planetary bodies. This process is rich in physics and was recently succinctly captured in Figure 1 of \citet{vanlie18:mnras480}. 

A complete understanding of the process detailed above is required to accurately tranlate the observed atmospheric abundances back into planetary abundances, and in this work we focus on the evolution of material through the compact accretion disk. The final stage of material transport through the accretion disk is poorly understood, but fortunately the accretion disks are accessible through direct observation. The disks are expected to host material in dusty and gaseous phases, both of which have been observed \citep{far16:nar71}. The warm dust emits primarily in the near-infrared radiation, and has been detected around several dozen white dwarf stars \citep{far16:nar71}. Signatures of gaseous debris have also been observed either in emission or absorption, but, despite the expectation that all systems with dusty debris should also host gaseous debris, observable gaseous debris appears to be a much rarer phenomena \citep{man16:mnras462}. 

The paucity of systems with observed gaseous debris in absorption is easily understood as a selection effect, requiring the gas to lie along the line-of-sight between the observer and the system. The observability of gaseous debris in emission should however be inclination independent. To date, only eight systems with both dusty debris and gaseous debris in emission have been confirmed \citep{man16:mnras462}, with no clear explanation as to why some dust systems host gaseous debris in emission while others do not. These systems show evidence of global evolution in timescales of years to decades \citep{man16:mnras462, den18:apj854}, and are extremely useful for testing theories of the accretion disk evolution \citep{boc11:apj741, mir18:apj857}.

In this project we explore methods for isolating the dusty debris disk systems which also host gaseous debris using only the photometric properties in an effort to understand their origin and provide a target selection method for future studies. In \cite{den17:apj849}, we introduced a method for selecting new dusty debris disk candidates based on the parameters of a single-temperature blackbody fit to the observed infrared excess. We also drew attention to the locus of gaseous debris disk hosting white dwarfs in the single-temperature blackbody plane, which forms the basis of this investigation. 

\section{Single-Temperature Blackbody Fits to Dusty Infrared Excesses}

The favored choice for modeling the contribution of the dust disk to the observed spectral energy distribution (SED) is the optically thick, geometrically thin model introduced by \citet{jur03:apjl584}. While extraordinarily successful, this model suffers from strong degeracies between the disk inclination and outer disk radius, making comparisons of literature determined parameters for different systems difficult. For this study, we focus instead on a simpler model of the infrared excess, a single-temperature blackbody, which allow direct comparison between systems while still capturing some of the information provided by the SED. 

\begin{figure}
  \centerline{\includegraphics[width=1.0\columnwidth]{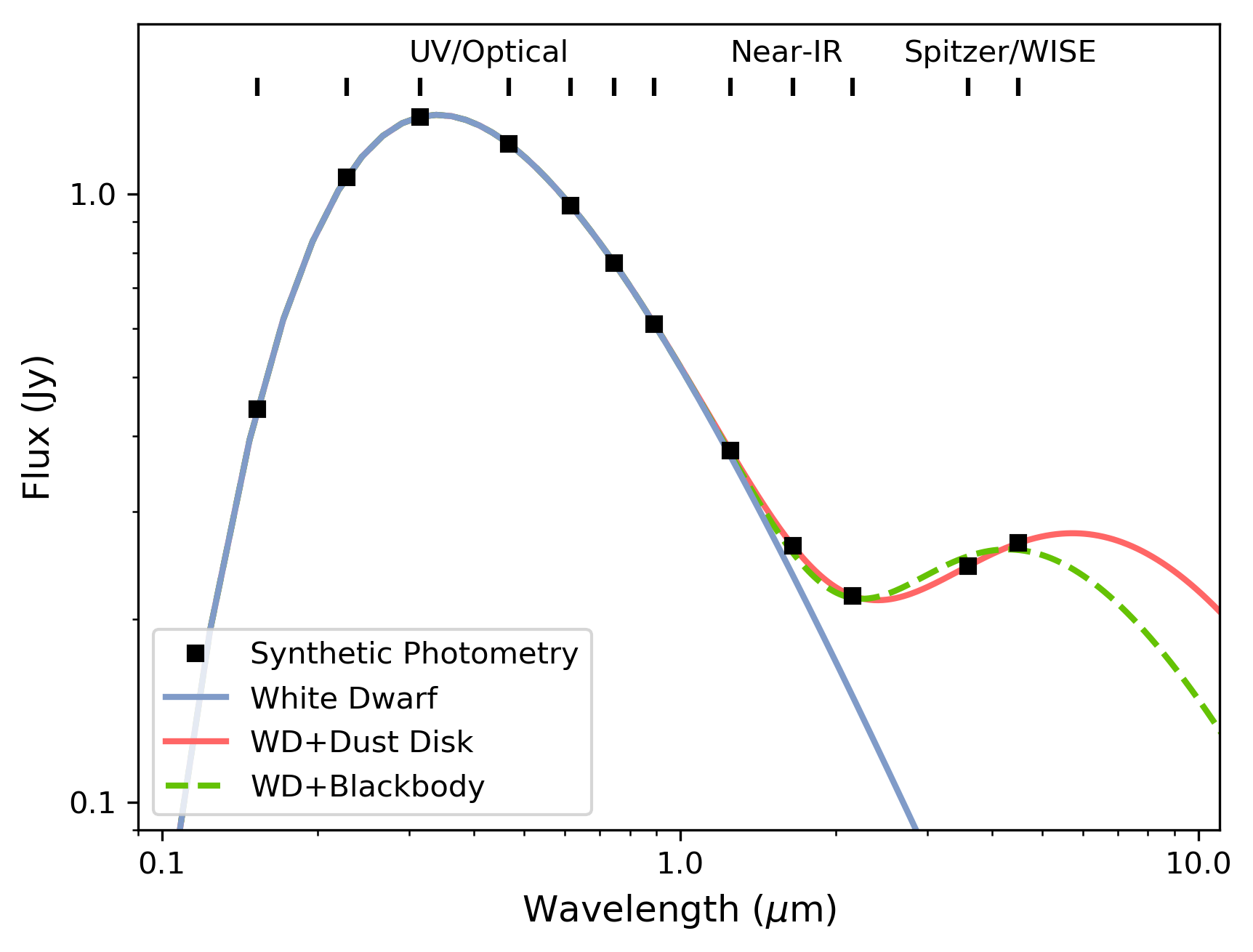}}
  \caption{A comparison between a single-temperature blackbody fit (dashed green) and an optically thin dust model fit (solid red) to simulated photometry. The dust infrared excesses perform reasonably well in the bands we typically observe, which extends to the \emph{Spitzer} IRAC Channels 1 and 2, or, similarly, the \emph{WISE} W1 and W2 bands.} 
  \label{sed}
\end{figure}

Before exploring the sample, it is worth confirming that the single-temperature blackbody is a reasonable model to apply. After all, the dusty debris disks are neither spherical nor are they composed of dust at a single-temperature, so it seems poorly motivated. But while the physical motivation is lacking, the single-temperature blackbody provides a reasonable estimate of the system flux in the bands in which we are interested, as shown in Fig.~\ref{sed}. Most observations of near-infrared excess from dusty debris disks are confined to wavelengths below 4.5 $\mu$m, where the \emph{WISE} all-sky survey is most sensitive and \emph{Spitzer} post-cryogenic photometry has continued to be available. 

The single-temperature blackbody does a reasonable job capturing the bulk properties of the dusty infrared excess. In a sense, one can think of the blackbody temperature as a flux-weighted average temperature of the disk. The hottest, inner dust dominates the fit, but this effect can be slightly offset by the radial extent of the disk, where cooler dust has a larger surface area. The blackbody radius on the other hand is a proxy for the projected surface area of the disk, dominated by the disk inclination. With this intuition in mind, we apply the single-temperature blackbody model to a well characterized sample of dusty white dwarfs. 

For our collection of dusty white dwarfs, we choose the \emph{Spitzer}-confirmed sample of \citep{roc15:mnras449}. This sample consists of 35 dust debris disk hosting white dwarfs with IRAC 3.6 and 4.5 $\mu$m band photometry, and was analyzed by the authors to search for trends in the fractional infrared luminosity as a function of white dwarf properties \citep{roc15:mnras449}. As their proxy for the infrared luminosity of the dust disk, the authors fit single-temperature blackbody models to the observed infrared excess, and we adopt their best-fitting parameters given in Table 3 for our study. 

With these in hand, we define the single-temperature blackbody plane with the two free parameters of the model, the temperature and radius of the blackbody. In Fig.~\ref{bbody_gas} we plot the best-fitting parameters of all 35 systems as grey diamonds, giving us our first look at the titular plane. Previously, we have used this plane to define a selection criteria for new dusty debris disk candidates \citep{den17:apj849}, but here, we focus only on the properties of previously discovered systems. 

\section{Gaseous Emission Boundary in the Single-Temperature Blackbody Plane}

\begin{figure}[h!]
  \centerline{\includegraphics[width=1.0\columnwidth]{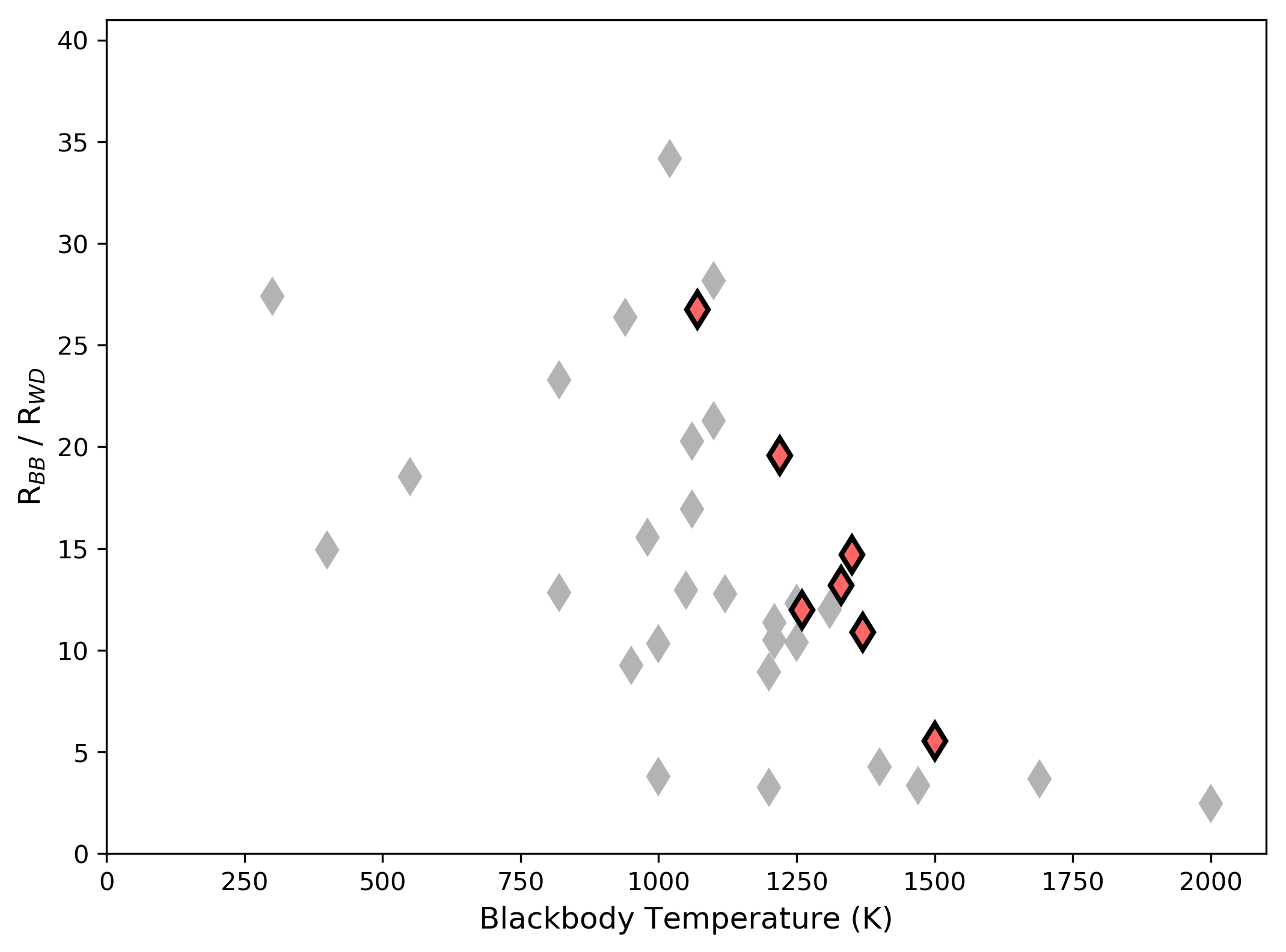}}
  \caption{The locations of all 35 dusty debris disks identified in \citet{roc15:mnras449} in the single-temperature blackbody plane as grey diamonds. We highlight in red the seven systems in this sample also known to host gaseous debris in emission, which aggregate at the high-temperature boundary of the dust disk region.} 
  \label{bbody_gas}
\end{figure}

There are a few key features of the region occupied by the dust disks we draw attention to before examining the gas-disk boundary. First, the majority of systems congregate between temperatures of 1000 and 1500 K. This follows the expectation that the dust should exist in the region bounded by the sublimation radius at the innermost edge and the tidal disruption radius at the outermost edge, which spans a temperature range of approx. 300-1800 K. Any dust within the sublimation radius will be sublimated into gas and dust production is expected to begin at the tidal disruption radius, beyond which larger bodies are expected to remain mostly intact. 

In accordance, the lack of observed systems at high temperatures (above 1500 K) reflects the sublimation of the dust. However, the dearth of systems at lower blackbody temperatures is potentially a selection effect, as these dust systems are fainter and only visible at longer wavelenghts where the data are limited. 

By identifying the systems which also host gaseous debris in emission as red diamonds on Fig.~\ref{bbody_gas} we quickly see that they congregate along the high-temperature boundary of the dust disk region. By itself, this is useful as a selection technique for future targeted studies of gaseous emission line hosting dust disks, but it is also interesting, and perhaps suggestive, in the context of dust disk evolution. 

For example, in the PR-driven evolutionary scenario explored by \citet{boc11:apj741}, as the outer edge of the disk evolves inward the average dust temperature increases, leading to higher best-fitting single-temperature blackbodies. This would lead to a dust disk evolving toward the boundary before eventually sublimating entirely into gas, suggesting that the gas-disk boundary could be an end-state of evolution for the dusty debris disks.

In another interpretation, collisions of bodies with existing dusty debris disks are expected to both produce copious amounts of gas and lead to temporary infrared brightening of the dust disks \citep{ken17:apj844, ken17:apj850}, which would lead to larger best fitting-radii, forcing systems towards the boundary from below. Neither of these interpretations is preferred by our analysis, and they are only given as examples of how the boundary could be interpreted. 

\section{Conclusions}

In the single-temperature blackbody plane the dusty debris disk systems which host gaseous debris in emission form the terminus of the dusty debris disk region, providing a useful selection criterion for targeted gas emission searches. This technique is well-suited for application to the expected large sample of \emph{GAIA} white dwarfs \citep{gen18:arxiv} with \emph{WISE} infrared excesses as it relies only on publicly available photometry for target selection.


\bibliography{wdexoplanets}


\end{document}